\magnification=\magstep1
\baselineskip=20pt
\centerline{Statistics and Microphysics of the Fracture of Glass}
\bigskip
\centerline{J. I. Katz}
\centerline{Department of Physics and McDonnell Center for the Space
Sciences}
\centerline{Washington University, St. Louis, Mo. 63130}
\centerline{katz@wuphys.wustl.edu}
\bigskip
\centerline{Abstract}
\medskip
The tensile strength of fused silica fibers is believed to approach its
intrinsic value at low temperature, and modern experiments indicate very
small, perhaps unmeasured, intrinsic dispersion in this strength.  I
consider the application of classical ``weakest link'' models to this
problem in an attempt to determine the number and therefore the nature of
the failure sites.  If the skewness as well as the dispersion (Weibull
modulus) of failure strengths are measured it may be possible to determine
both the number of sites and the distribution of their strengths.  Extant
data are not sufficient, but I present calculated skewnesses for comparison
with future data.
\vfil
\eject
\noindent
{\bf I INTRODUCTION}
\medskip
Fused silica optical fiber is one of very few materials whose tensile
strength, measured at low temperature (77$^\circ$K), is believed to equal
its theoretical limiting strength$^{1,2}$.  Failure occurs at strains of
about 0.2, extraordinarily large for a brittle material, but consistent with
the idea that its strength is limited only by the strength of its
constituent molecular bonds.  In addition, it fails under accurately
reproducible loads, which is very unusual for a brittle material.  These
properties may be explained by the absence, in properly prepared samples, of
surface cracks which would concentrate stress.

The remarkable properties of glass fiber, if carefully manufactured and
protected from abrasion by a soft coating, have been recognized for several
decades$^{3,4}$.  Most research has been concerned with the fracture of
glass under tension at temperatures comparable to room temperature, at which
it is subject to chemical attack by water vapor or liquid and shows static
fatigue$^5$.  However, at temperatures around 77$^\circ$K activation
barriers are believed to reduce the rate of chemical attack to a negligible
level, and the ultimate material strength may be observed.

The results of early experiments$^4$ at 78$^\circ$K found scattered tensile
strengths of up to 130 Kbar.  Later experiments$^6$ on fused silica measured
tensile strengths as high as 150 Kbar, but also showed substantial scatter;
these authors found no evidence for a significant difference between the
strength measured at 77$^\circ$K and that at 4$^\circ$K.  More recent
experiments$^{7,8}$ on fused silica at 77$^\circ$K reported very little
scatter and inferred a limiting tensile strain in the range 0.18--0.21;
using the zero-stress value of Young's modulus this implies a limiting
stress of $\approx 140$ Kbar.

Despite the internal consistency of the data$^{7,8}$ the absolute strain at
failure is not known quantitatively, because this experiment used a
two-point bending apparatus, whose analysis depends on a theory which
assumes a slender fiber, small strain, and linear stress-strain relation,
none of which are quantitatively valid at these values of strain.

Silica-based glasses with compositions other than pure SiO$_2$ qualitatively
resemble fused silica, but are typically somewhat weaker$^9$.  At higher
temperatures chemical attack makes the strength of glass depend on the
temperature, chemical environment (humidity and pH), and the duration of
loading (static fatigue is evident), and measured tensile strengths are
typically a factor of three less than at 77$^\circ$K$^{4-13}$.

The reproducibility of measurements of strength is usually described by the
Weibull modulus$^{14,15}$ $m$; the fractional dispersion in measurements of
failure load is $\approx [m(m-1)]^{-1/2}$.  Most engineering ceramics$^{15}$
have $m \approx 10$, and glass with surfaces abraded deliberately or by
ordinary handling$^{16}$ has $m$ in the range 3--7.  Measured$^{7-11,13}$
Weibull moduli for modern fused silica optical communications fiber are in
the range 30--100, both at low temperatures and around room temperature.  It
has been argued$^{17}$ that even this small apparent dispersion is the
consequence of dispersion in fiber diameter (and hence in stress at a given
applied force) rather than intrinsic variation in the properties of the
glass.

The distribution of failure stresses may contain useful information about
the mechanism of failure.  For example, brittle materials, including glass,
with abraded or damaged surfaces have empirical tensile strengths far below
their ideal strengths.  They also have small Weibull moduli.  These facts
are the consequence of stress concentration at flaws$^3$, usually surface
scratches; the inevitable variation from sample to sample in the size and
shape of these flaws produces variation in measured sample strength and
gives a small Weibull modulus.

Even the best glass fibers should show some level of sample-to-sample
variation.  Because glass is amorphous its chemical bonds are each in a
subtly different environment, and some should be weaker, or strained more by
a macroscopic strain field, than others.  In addition, real silica glass is
not perfectly pure SiO$_2$ (or Si$_x$Ge$_{1-x}$O$_2$ in an optical fiber),
but contains chemical impurities which may locally weaken it.

The purpose of this paper is to develop simple statistical models of the
dependence of the distribution of failure strengths on the number of
independent elements $N$ whose weakest member initiates failure.  If the
distribution of material strength is measured, and the basic parameters of
the model are known, it will then be possible to estimate $N$.  The value of
$N$ constrains the nature of the sites at which failure is initiated: if $N$
is comparable to the number of atoms in the sample these must be ordinary
Si---O bonds; if $N$ is of order the number of transition metal impurity
atoms then these are implicated, while a small $N$ (in the range 10--10$^4$,
for example) would implicate macroscopic flaws or heterogeneities, such as
the surface scratches of inclusions which are responsible for the low
strength and low Weibull modulus of ordinary glass.  In the case of
room-temperature failure, surface sites for chemical attack also need be
considered.

Statistical ``weakest link'' theories of fracture are not new$^{18,19}$.
What is new, since the foundations of this subject were laid, is the
development of optical communications fiber as a mass produced material
whose strength, at least at low temperature, approaches its ideal value
because macroscopic samples may be {\it completely} free of
stress-concentrating flaws of greater than atomic size.  Hence its fracture
statistics contain information about the microphysics of its ideal strength.
\bigskip
\noindent
{\bf II DISPERSION OF FAILURE STRESS}
\medskip
In ``weakest link'' failure models a sample is assumed to consist of $N$
sites, or links, each of which is described by a fracture readiness
parameter $\alpha$, which may be considered a stress concentration factor,
or the reciprocal of a link's ideal strength.  The values of $\alpha$ are
distributed according to a distribution $f(\alpha)$, which is defined for
$\alpha > 0$ and normalized
$$\int^\infty_0 f(\alpha^\prime)\,d\alpha^\prime = 1. \eqno(1)$$
The sample fails if the largest fracture readiness parameter found among the
$N$ sites, $\alpha_{max}$, exceeds a value $\alpha_0$.  The probability that
it does not fail is, to good approximation if $N \gg 1$,
$${\cal P}(\alpha_0) \approx \exp\left(-\int^\infty_{\alpha_0} N
f(\alpha^\prime)\,d\alpha^\prime\right). \eqno(2)$$
The probability that failure occurs for a value of $\alpha_0$ between
$\alpha$ and $\alpha + d\alpha$ is $P(\alpha)\,d\alpha$, where
$$P(\alpha) = Nf(\alpha)\exp\left(-\int^\infty_\alpha
Nf(\alpha^\prime)\,d\alpha^\prime\right). \eqno(3)$$

Unfortunately, we have no {\it a priori} knowledge of the functional form
of $f(\alpha)$.  I therefore consider several possible forms.  The most
useful way to parametrize the results is as the ratio of the width $w$
of $P(\alpha)$ to the value $\alpha_{max}$ at which $P(\alpha)$ is a
maximum; in terms of the Weibull modulus
$${w \over \alpha_{max}} \approx {1 \over [m(m-1)]^{1/2}}, \eqno(4)$$
and the approximation is almost exact if $w$ is defined as the dispersion of
$P(\alpha)$
$$w \equiv \left(\left\vert{d^2\ln P(\alpha) \over
d\alpha^2}\right\vert_{\alpha=\alpha_{max}}\right)^{-1/2} \eqno(5)$$
and $m \gg 1$.

A ``boxcar distribution'' is defined
$$f(\alpha) = \cases{1/\alpha_b,&if $0 \le \alpha \le \alpha_b$;\cr 0,&if
$\alpha > \alpha_b$.\cr} \eqno(6)$$
Here $\alpha_{max}=\alpha_b$.  The full width at half maximum of $P(\alpha)$
$w^\prime = \alpha_{max} \ln 2/N$ and the fractional width is
$${w^\prime \over \alpha_{max}} = {\ln 2 \over N}. \eqno(7)$$
This distribution implies unmeasurably small $m$, but is unrealistic {\it a
priori} (why should the distribution of fracture readiness drop
discontinuously from a constant value to zero?).  It also disagrees with
available data$^{7-9}$ on glass at low temperatures where its intrinsic
strength appears to be acheived (as well as with all other data on brittle
materials).

An exponential distribution is defined
$$f(\alpha) = {1 \over \alpha_0} \exp(-\alpha/\alpha_0). \eqno(8)$$
Then
$${w \over \alpha_{max}} = {1 \over \ln N}. \eqno(9)$$
If $f(\alpha)$ is exponential then a useful estimate of $N$ may be possible
because $w/\alpha_{max}$ should be measurable and is sensitive enough to $N$
to constrain it significantly.

A power law distribution with $p > 1$ is defined
$$f(\alpha) = \cases{(p-1) \alpha_0^{p-1} \alpha^{-p},&if $0 < \alpha_0 \le
\alpha$;\cr 0.&if $\alpha < \alpha_0$,\cr} \eqno(10)$$
Then
$${w \over \alpha_{max}} = {1 \over [p(p-1)]^{1/2}}. \eqno(11)$$
The power law exponent $p$ equals the Weibull modulus $m$ and is independent
of $N$.  If $f(\alpha)$ is a power law then measurements of $m$ provide no
information about $N$.  This is implausible (as well as disappointing)
because the very large reported values of $m$ would imply implausibly steep
distributions of $\alpha$.

A Gaussian distribution is defined
$$f(\alpha) = {2 \over \alpha_0 \pi^{1/2}} \exp\left[-\left({\alpha \over
\alpha_0}\right)^2\right]. \eqno(12)$$
For this distribution $\alpha_{max}$ and $w$ are found by successive
approximation.  The result is
$$\eqalign{{w \over \alpha_{max}}&\approx \left\{2\left[1 + 2\ln N^\prime
\left(1 - {\ln\ln N^\prime \over 2 \ln N^\prime}\right)^{1/2}\right]
\left(\ln N^\prime - {1 \over 2}\ln\ln N^\prime\right)\right\}^{-1/2}\cr
&\approx {1 \over 2 \ln N^\prime},\cr}\eqno(13)$$
where $N^\prime \equiv N/\pi^{1/2}$ and the simple approximation in the
second line of Eq. 13 is usually accurate to better than 10\%.  $P(\alpha)$
is shown in Fig. 1 for several values of $N$ of interest.

The stretched exponential function is defined
$$f(\alpha) = {C(\nu) \over \alpha_0}\exp\left[-\left({\alpha \over
\alpha_0}\right)^\nu\right]. \eqno(14)$$
This is a general form widely used when the actual functional form is
unknown, and includes the simple exponential and Gaussian as special cases.
The normalizing constant $C(\nu) \equiv \nu/\Gamma(1/\nu)$ and $N^\prime
\equiv NC(\nu)/\nu$.  By successive approximations
$$\eqalign{{w \over \alpha_{max}}&\approx {\left\{1 - {1 \over 2 \ln
N^\prime}\left[{\nu-1 \over \nu} - \left({\nu-1 \over \nu}\right)^2\ln\ln
N^\prime\right]\right\} \over \nu \ln N^\prime\left(1 - {\nu-1 \over \nu}
{\ln\ln N^\prime \over \ln N^\prime}\right)^{1/\nu}}\cr &\approx {1 \over
\nu\ln N^\prime}.\cr} \eqno(15)$$

It is evident that if $f(\alpha)$ is, or can be fitted to, a stretched
exponential or to one of its special cases useful and plausible estimates of
$N \approx \exp(\alpha_{max}/(w\nu))$ may be obtained.  However, the
inferred value of $N$ depends very sensitively on $\nu$, and measurement of
$m$ alone for a sample of test objects of a single size does not determine
$\nu$.  Measurement of two or more populations of very different-sized test
objects of the same material (for which $N$ is proportional to the size) may
determine both $N$ and $\nu$, and may be feasible; for example, in two-point
bending experiments$^{7-9}$ on optical fiber of 125$\mu$ diameter the number
of atoms $N_a$ at significant risk of initiating fracture (the fraction with
stresses within about $1/m$ of the maximum, which are found only close to
the outside of the sharpest part of the bend) is $N_a \approx 5 \times
10^{14}$, while in tensile loading of a 50 m gauge length$^{10}$ of fiber
$N_a \approx 4 \times 10^{22}$ atoms are uniformly stressed.  Two or more
measurements of $m$ in which very different numbers of atoms are stressed
permit simultaneous determination of both the ratio $N/N_a$ and $\nu$,
although no extant data serve the purpose.
\bigskip
\noindent
{\bf III SKEWNESS}
\medskip
A possible method of determining $\nu$, and hence $N$, is to measure the
skewness of $P(\alpha)$:
$$s \equiv {\int^\infty_0 (\alpha-\alpha_{max})^3 P(\alpha)\,d\alpha \over 
w^3 \int^\infty_0 P(\alpha)\,d\alpha}. \eqno(16)$$
The skewness is not small; see Fig. 1.  It is also not readily estimated
analytically because the Taylor expansion of $P(\alpha)$ about
$\alpha_{max}$ does not converge sufficiently rapidly, but it may be
calculated numerically.  Values of the skewness, as a function of $\nu$ and
$N$, are shown in Fig. 2.

The values plotted were computed using cutoffs on the integrals of $\pm 5w$
from $\alpha_{max}$, with $w$ defined self-consistently using the same
cutoff.  The reason for this is that $P(\alpha)$ has a long tail toward
increasing $\alpha$ which contributes significantly to the skewness, but
which is unlikely to be observed in a real experiment with a reasonable
number of samples because there will probably be no samples that far out in
the tail.  The skewness computed without this cutoff is significantly
larger, typically by $O(10\%)$.

When comparing experimental data to Fig. 2 a similar cutoff must be applied
to the data.  This will, in addition, exclude samples which are anomalously
weak because of mechanical damage or other gross flaws, which otherwise must
be excluded {\it ad hoc}.  If the number of measurements is not large it may
be necessary to choose a narrower cutoff, and to recompute Fig. 2
accordingly.
\bigskip
\noindent
{\bf IV DISCUSSION}
\medskip
It is generally assumed that the observed strength of pristine glass fibers
at low temperatures (and possibly also at higher temperatures in an inert
environment) is the intrinsic strength of the material.   However, this has
not been proved, and may not be so.  For example, the strength could be
limited by the presence of trace impurities, in which case their reduction
or elimination would increase the strength.

The actual intrinsic strength (or, equivalently, the limiting strain at
failure) of fused silica is not accurately known.  As discussed above,
obtaining the intrinsic strength from two-point bending experiments$^{7-9}$
requires use of the nonlinear stress-strain relation, which is not known to
the required strain level, in a three-dimensional elastostatic calculation.
Nonlinear stress-strain data$^{4,20}$ extend up to strains of about 0.16,
but show the Young's modulus still increasing with strain.  This sense of
nonlinearity is the same as that found for rubbery elastomers, in which
polymer chains are easily straightened at small stress, but which become
much stiffer when stretched to their full length; the analogous effect in
silica glass involves the comparatively soft bending of the O---Si---O bond 
angles towards 180$^\circ$, followed by the greater resistance of the bonds
to extension.  However, in elementary models$^{21}$ of zero temperature
ideal strength the effective Young's modulus must decrease to zero at the
point of failure, which corresponds to the inflection point of the
interatomic potential; there is as yet no sign of this decrease in the data.
This is qualitatively consistent with the reported$^{7,8}$ limiting strains
of 0.18--0.21, but the uncertainty of this value and the lack of modern data
on the stress-strain relation at high strain call for further experiments.

It should be considered whether the observed reduction in strength of
optical fiber at room temperature may in part be the consequence of thermal
excitation, reducing the mechanical work required to disrupt a bond.  In
order of magnitude this is plausible: the greatest reduction in required
mechanical work per bond is $\approx k_B T \ln N_t \approx 80 k_B T$, where
the number of trials of bonds $N_t \approx N_b t/\omega$, where $N_b$ is
the number of bonds in the sample, $t$ the duration of the experiment and
$\omega$ a typical vibration (or vibrational relaxation) frequency.
Evaluating at room temperature and dividing by the volume per bond yields a
reduction in limiting stress $\approx 100$ Kbar, approximately equal to the
measured reduction in failure stress between 77$^\circ$K and room
temperature.  However, a proper evaluation of thermally assisted fracture
takes the form of nucleation theory$^{22}$, which predicts a temperature
dependence of the strength quite different from the available data$^6$ (only
vacuum data are relevant).  In addition, spontaneous thermal nucleation of
fracture is not characterized by a waiting time.  Such a waiting time is
observed$^{9,11,12}$ in experiments in hostile chemical environments, but
it is not known if there is a waiting time in an inert environment.

Most of the extant data on the failure of glass fibers were obtained in warm
humid or wet environments in which glass is subject to corrosion by water.
Neither the methods of this paper nor nucleation theory are directly
applicable to failure by stress corrosion, but these data do show large
(typically in the range 30--100) Weibull moduli.  It may be that a modified
version of weakest link theories, in which the parameter $\alpha$ represents
reactivity rather than stress concentration or mechanical bond strength, may
be applicable.  In that case the inferred value of $N$ would provide
information about the number and kind of surface sites vulnerable to
chemical attack.  

I thank DARPA for support and M. C. Ogilvie for assistance with computer
graphics.
\vfil
\eject
\centerline{References}
\bigskip
\item{$^1$} C. R. Kurkjian and D. Inniss {\it Opt. Eng.} {\bf 30}, 681
(1991).
\item{$^2$} C. R. Kurkjian, P. G. Simpkins and D. Inniss {\it J. Am. Ceram.
Soc.} {\bf 76}, 1106 (1993).
\item{$^3$} A. A. Griffith, {\it Phil. Trans. Roy. Soc.} {\bf A221}, 163
(1920).
\item{$^4$} W. B. Hillig, {\it Modern Aspects of the Vitreous State} V. 2,
ed. J. D. Mackenzie (Butterworths, Washington) p. 152 (1962).
\item{$^5$} R. J. Charles, {\it J. Appl. Phys.} {\bf 29}, 1554 (1958).
\item{$^6$} B. A. Proctor, I. Whitney and J. W. Johnson, {\it Proc. Roy.
Soc. (London)} {\bf 297A}, 534 (1967).
\item{$^7$} P. W. France, M. J. Paradine, M. H. Reeve and G. R. Newns {\it
J. Mat. Sci.} {\bf 15}, 825 (1980).
\item{$^8$} W. J. Duncan, P. W. France and S. P. Craig {\it Strength of
Inorganic Glass} ed. C. R. Kurkjian (Plenum, New York) p. 309 (1986).
\item{$^9$} P. W. France, W. J. Duncan, D. J. Smith and K. J. Beales {\it J.
Mat. Sci.} {\bf 18}, 785 (1983).
\item{$^{10}$} C. R. Kurkjian, R. V. Albarino, J. T. Krause, H. N. Vazirani, F.
V. DiMarcello, S. Torza and H. Schonhorn {\it Appl. Phys. Lett.} {\bf 28},
588 (1976).
\item{$^{11}$} H. C. Chandan and D. Kalish {\it J. Am. Ceram. Soc.} {\bf 65},
171 (1982).
\item{$^{12}$} M. J. Matthewson and C. R. Kurkjian {\it J. Am. Ceram. Soc.}
{\bf 71}, 177 (1988).
\item{$^{13}$} J. E. Ritter, T. H. Service and K. Jakus {\it J. Am. Ceram.
Soc.} {\bf 71}, 988 (1988).
\item{$^{14}$} L. G. Johnson {\it The Statistical Treatment of Fatigue
Experiments} (Elsevier, Amsterdam) (1964).
\item{$^{15}$} M. F. Ashby and D. R. H. Jones {\it Engineering Materials 2:
An Introduction to Microstructures, Processing and Design} (Pergamon,
Oxford) (1986).
\item{$^{16}$} J. E. Ritter, Jr. {\it J. Appl. Phys.} {\bf 40}, 340 (1969).
\item{$^{17}$} C. R. Kurkjian and U. C. Paek {\it Appl. Phys. Lett.} {\bf
42}, 251 (1983).
\item{$^{18}$} B. Epstein {\it J. Appl. Phys.} {\bf 19}, 140 (1948).
\item{$^{19}$} R. H. Doremus {\it J. Appl. Phys.} {\bf 54}, 193 (1983).
\item{$^{20}$} J. T. Krause, L. R. Testardi and R. N. Thurston {\it Phys.
Chem. Glasses} {\bf 20}, 135 (1979).
\item{$^{21}$} I. N\'aray-Szab\'o and J. Ladik {\it Nature} {\bf 188}, 226
(1960).
\item{$^{22}$} J. C. Fisher {\it J. Appl. Phys.} {\bf 19}, 1062 (1948).
\vfil
\eject
\centerline{Figure Captions}
\bigskip
\noindent
Figure 1: Distribution $P(\alpha)$, normalized to $P(\alpha_{max})$, as a
function of $(\alpha-\alpha_{max})/w$ for a Gaussian $f(\alpha)$.  The solid
curve is for $N = 10^2$, short dashed curve for $N = 10^5$ and long dashed
curve for $N = 10^{20}$ ($N = 10^{15}$ is indistinguishable from $N =
10^{20}$ at the resolution of the figure, but these values may be
distinguished by their different values of $w/\alpha_{max}$ and predicted
$m$).  The substantial skewness is evident.
\medskip
\noindent
Figure 2: Contour graph of calculates skewness of $P(\alpha)$ as a function
of $\nu$ and $N$.  Integrations (to calculate both skewness and dispersion)
were self-consistently truncated at $\vert \alpha - \alpha_{max}\vert = 5w$.
\vfil
\eject
\end